\documentclass{JHEP3}
\usepackage{amsmath}
\DeclareMathOperator{\Tr}{Tr}
\newcommand{\ospsixtwo}{\text{osp}^*(8|4)}
\newcommand{\eqn}[1]{(\ref{#1})}  
\title{Anti-de-Sitter vacua require fermionic brane charges}
\preprint{hep-th/0311110\\AEI-2003-092}
\author{Kasper Peeters and Marija Zamaklar\\
MPI/AEI f\"ur Gravitationsphysik\\
Am M\"uhlenberg 1\\
14476 Golm, GERMANY\\
E-mail: \email{kasper.peeters, marija.zamaklar@aei.mpg.de}}

\abstract{We argue that a modification of the super-AdS algebras which
  accounts for the presence of D-branes requires not only the
  inclusion of \emph{bosonic} brane charges, but also the inclusion of
  new \emph{fermionic} ones. We show that such fermionic brane charges
  are indeed present in the matrix model and the supermembrane in the
  pp-wave limit of the corresponding backgrounds.  We briefly comment
  on an AdS version of Sezgin's M-algebra inspired by this
  observation.}

\begin{document}
\section{Introduction}

Rather surprisingly, a modification of the superalgebra of
anti-de-Sitter backgrounds\footnote{When we talk about anti-de-Sitter
backgrounds in this letter, we always mean the maximally
supersymmetric $\text{AdS}_p\times S^q$ backgrounds. Similarly, we use
the phrase ``anti-de-Sitter (super)algebras'' as a shorthand to refer
to the (super)isometry algebras of these product backgrounds.} which
accounts for the presence of D-branes in the string spectrum is still
unknown. At an algebraic level, D-branes manifest themselves through
non-zero expectation values of bosonic tensorial charges. There exists
a widespread, but incorrect, belief that the inclusion of these brane
charges into the anti-de-Sitter superalgebras follows the well-known
flat-space pattern. In flat space, the inclusion of brane charges
leads to a rather minimal modification of the super-Poincar\'e
algebra: the bosonic tensorial charges appear on the right-hand side
of the anti-commutator of supercharges, transform as tensors under the
Lorentz boosts and rotations, while they commute with all other
generators~\cite{town3}.  The brane charges are therefore often
loosely called \mbox{``central''}, and the resulting algebra is
referred to as the maximal bosonic \mbox{``central''} extension of the
super-Poincar\'e algebra. However, despite several attempts to
construct a similar modification of anti-de-Sitter superalgebras, a
\emph{physically satisfactory} solution is as of yet unknown.

There are two basic physical requirements which have to be satisfied
by an anti-de-Sitter algebra which is modified to include brane
charges. The algebra has to include at least the brane charges which
correspond to all D-branes that are already known to exist, and it
also has to admit at least the supergraviton multiplet in its
spectrum.  Mathematically consistent modifications of anti-de-Sitter
superalgebras can be constructed, but all existing proposals fail to
satisfy one or both of these physical criteria. We refer the reader
to~\cite{Meessen:2003yi} for an extensive discussion of this problem
and the existing literature on this topic. In the present letter we
show that there is a simple reason why previous attempts to extend
anti-de-Sitter superalgebras with brane charges have failed: such
extensions are \emph{only} physically acceptable when one adds new
\emph{fermionic} brane charges as well.

The necessity of including new fermionic brane charges into the
modified algebra can be understood from a very simple argument based
on Jacobi identities, in combination with the two physical
requirements just mentioned. Consider an anti-de-Sitter superisometry
algebra, or a pp-wave contraction of it. The bracket of supercharges
can, very symbolically, be written in the form
\begin{equation}
\{ Q_\alpha, Q_\beta \} = (\Gamma^{AB})_{\alpha\beta} M_{AB}\, ,
\end{equation}
where $Q$ and $M$ are the supercharges and rotation generators
respectively (we have grouped together momentum and rotation
generators by using a notation in the embedding space). Suppose now
that we add a bosonic tensorial brane charge~$Z$ on the right-hand
side of this bracket. This extension has to be made consistently with
the Jacobi identities. Consider the $(Q,Q,Z)$ identity, which takes
the symbolic form
\begin{equation}
\label{e:QQZ}
\begin{aligned}
( Q_\alpha, Q_\beta, Z ) 
&=  [\{Q_\alpha,Q_\beta\}, Z] 
- \big[[Q_\alpha, Z], Q_\beta\big]
- \big[[Q_\beta, Z], Q_\alpha\big]  \\[1ex]
&= (\Gamma^{AB})_{\alpha\beta} [ M_{AB}, Z]
 -2\,\big[[Q_{(\alpha}, Z],Q_{\beta)}\big]\,.
\end{aligned}
\end{equation}
As the brane charge~$Z$ is a tensor charge, it will transform
non-trivially under the rotation generators. This implies that the
first term of~\eqn{e:QQZ} will not vanish. The Jacobi identity can
then only hold if~$Z$ \emph{also} transforms non-trivially under the action
of the supersymmetry generators! (In flat space, only the
vanishing bracket $[P,Z]$ appears in the first term of the Jacobi
identity~\eqn{e:QQZ}, because in that case the $\{Q,Q\}$
anti-commutator closes on the translation generators).  The simplest
option is to assume that \emph{no new fermionic charges} should be
introduced, and that therefore symbolically
\begin{equation}
\label{e:QZisQ}
[ Q_\alpha, Z ] = Q_\alpha \, .
\end{equation}
Although it is possible to construct an algebra based on~\eqn{e:QZisQ}
which satisfies all Jacobi identities, it is physically
unsatisfactory~\cite{Meessen:2003yi}. The essential reason is that
brackets like~\eqn{e:QZisQ} are incompatible with multiplets on which
the brane charge is zero (the left-hand side would vanish on all
states in the multiplet, while the right-hand side is not zero).  In
other words, one cannot ``turn off'' the brane charges.\footnote{The
precise situation is a slightly complicated since there is more than
one bosonic brane charge; it could in principle be that a subtle
interplay between these charges resolves the problem mentioned
above. The full analysis of~\cite{Meessen:2003yi} shows that this
mechanism can, however, not be realised. Similarly, we have shown that
it is not possible to close the Jacobi identities by e.g.~assuming
non-standard transformation behaviour of the brane charges under the
rotation generators. \cite{Meessen:2003yi} contains a full list of
(failing) alternatives.} The only other way out is to add \emph{new
fermionic charges}~$Q'_\alpha$ to the algebra, such that~\eqn{e:QZisQ}
is replaced with
\begin{equation}
[ Q_\alpha, Z ] = Q'_\alpha \, .
\end{equation}
In this case it becomes possible to find representations in which both
$Z$ and the new charge $Q'_\alpha$ are realised trivially, as expected
for e.g.~the supergraviton multiplet, while still allowing for
multiplets with non-zero brane charges.

This formal argument based on Jacobi identities may come as a
surprise, and one would perhaps find it more convincing to see new
fermionic brane charges appear in \mbox{\emph{explicit}} models. In
the present letter, we will show that such charges indeed do
appear. In order to show this, we will analyse the world-volume
superalgebras of the supermatrix model and the supermembrane in a
pp-wave limit of the anti-de-Sitter background. These models exhibit,
in the absence of brane charges, a world-volume version of the
superisometry algebra of the background geometry. When bosonic winding
charges are included, the algebra automatically exhibits fermionic
winding charges as well. Moreover, configurations on which these
charges are non-zero can be found explicitly, or can alternatively be
generated from configurations on which the fermionic winding charges
are zero. On the basis of these results we will briefly discuss a
D-brane extension of the $\ospsixtwo$ superisometry algebra with
bosonic as well as fermionic brane charges, which avoids the problems
with purely bosonic modifications as first observed
in~\cite{Meessen:2003yi}.

For historical completeness, we should mention that the extension of
world-volume superalgebras with fermionic central charges is not a new
idea. Green~\cite{Green:1989nn} has suggested a string world-volume
algebra in which a fermionic central charge is introduced as the
fermionic partner of the momentum generator. Generalisations of this
algebra to other brane world-volume superalgebras were made by Bergshoeff
and Sezgin~\cite{Bergshoeff:1995hm} and several other
authors. A~similar idea was implemented at the level of the
super-Poincar\'e isometry algebra of the Minkowski
vacuum~\cite{Bergshoeff:1989ax,Sezgin:1997cj}. In all of these
constructions, however, the extra fermionic brane or winding charges
were \emph{optional} rather than required for physical consistency.

\section{Fermionic brane charges}
\subsection{Matrix-model charges in a pp-wave}
\label{s:ppwave}

The supersymmetry algebra of the matrix model in a pp-wave background
contains brane charges. These charges (but not the full algebra!) have
been determined by Hyun and Shin~\cite{Hyun:2002cm} by computing the
Dirac bracket of the supercharges (see also the work of Sugiyama and
Yoshida~\cite{Sugiyama:2002rs} in which a similar calculation was done
for the supermembrane). In this section we will show that there are
further, fermionic brane charges, as expected from the Jacobi identity
argument sketched in the introduction.

Let us first briefly review the existing calculation as given
in~\cite{Hyun:2002cm}. In the matrix model, brane charges are ``traces
of commutators'', which identically vanish for finite~$N$. These
correspond to topological or winding charges (total derivatives) in
the supermembrane model. For example, the supermembrane has a two-form
winding charge, which is related to a charge in the matrix model
according to
\begin{equation}
Z^{IJ} = \int\!{\rm d}^2\sigma\, \epsilon^{0rs}\partial_r X^I \partial_s X^J
\quad\leftrightarrow\quad
{}Z^{IJ} = \Tr [X^I, X^J]\, .
\end{equation}
The ``integration over the world-volume'' in the supermembrane
corresponds to ``taking traces over $SU(N)$ indices'' in the matrix
model, and the trace over the commutator clearly only makes sense in
the $N\rightarrow\infty$ limit. In practise, one computes the algebra
of charge densities, or in matrix-model language, quantities obtained
before taking traces. See \cite{Hyun:2002cm} for more details, as well
as the papers by Banks et~al.~\cite{bank2} and Ezawa
et~al.~\cite{ezaw2}, where central charges were first computed in
matrix model context (though in flat space-time).

The action of M(atrix) theory in a pp-wave background is given by
\footnote{\label{f:conventions} We are using the conventions of
  Banks~et al.~\cite{bank2} and Hyun and Shin~\cite{Hyun:2002cm}. For
  the SO(9) gamma matrices this means that products satisfy the
  symmetry properties $\delta_{(\alpha\beta)},
  \gamma^I_{(\alpha\beta)}, \gamma^{IK}_{[\alpha\beta]},
  \gamma^{IKL}_{[\alpha\beta]}$ and so on. The 9-dimensional indices
  $I,K,L$ split in SO(3) indices $i,j,k$ and SO(6) indices
  $i',j',k'$. We also use the standard pp-wave symbol $\Pi =
  \gamma^{123}$.}
\begin{multline}
S = \Tr\bigg( \frac{1}{2R} D_0 X^I D_0 X^I
 -\frac{1}{2R} \left( \frac{\mu}{3} \right)^2 (X^i)^2
 -\frac{1}{2R} \left( \frac{\mu}{6} \right)^2 (X^{i'})^2
 + \frac{i}{2} \theta^\alpha D_0 \theta^\alpha\\[1ex]
 - \frac{i\mu}{8} \theta^\alpha
        \Pi_{\alpha \beta} \theta^\beta 
   + \frac{R}{4} [ X^I, X^J ]^2
 + \frac{R}{2} \theta^\alpha
      \gamma^I_{\alpha\beta}
      [ X^I, \theta^\beta ]
 - \frac{i\mu}{3} \epsilon_{ijk} X^i X^j X^k \bigg)\,
\end{multline}
Note that all fields $X,P$ and $\theta$ are~$SU(N)$ matrix valued and
therefore the ordering is important. The parameter $\mu$ is the pp-wave 
mass parameter and $R$ corresponds to the DLCQ radius; flat space is recovered
by taking $\mu\rightarrow 0$. The derivative~$D_0$ includes the usual
coupling to the world-line gauge field $D_0 X^I = \partial_0 - [
  \omega, X^I ]$; see~\cite{wit11} for further details.

The Hamiltonian, rotation generators, supercharges and several
bosonic non-per\-tur\-bative charges have been computed
in~\cite{Hyun:2002cm} and we will not comment any further on their
derivation. For reference, an overview of these charges (or rather
their ``un-integrated'' forms, i.e.~the expressions obtained before
taking the trace) is given in table~\ref{t:Hcharges}. We refer
to~\cite{kas_supwind} for a discussion of the $z^I$ charge.
\begin{table}[t]
\begin{subequations}
\begin{align}
\intertext{\bf Supercharges:}
q_{\alpha} &= \sqrt{\frac{R}{2}}\Big\{
   P^{I} \gamma^I_{\alpha\beta} 
 -\frac{i}{2}[X^I, X^J]\gamma^{IJ}_{\alpha\beta}
 -\frac{\mu}{3R} X^i (\Pi \gamma^i)_{\alpha\beta}
 +\frac{\mu}{6R} X^{i'} (\Pi \gamma^{i'})_{\alpha\beta},
   \,\theta^\beta \Big\}\,,\\
\label{e:Qkin}
\tilde q_{\alpha} &= \sqrt{\frac{2}{R}} \theta_{\alpha}\,,
\intertext{\bf Rotation generators:}
j^{ij} &= X^i P^j - P^i X^j - \tfrac{i}{4}\theta\gamma^{ij}\theta\,,\\[1ex]
j^{i'j'} &= X^{i'} P^{j'} - P^{i'} X^{j'} -
  \tfrac{i}{4} \theta\gamma^{i'j'}\theta\,,\\
\intertext{\bf Bosonic brane charges:}
z^I      &= i R\, \Big\{ P^J, \big[ X^J, X^I \big]\Big\}
 - \frac{R}{2} \Big[ \theta^{\alpha'}, \big[ \theta^{\alpha'}, X^I \big]\Big]\,,\\[1ex]
z^{IJ}   &= \frac{i}{2} \big[ X^I, X^J \big]\,,\\[1ex]
z^{IJKL} &= R\, X^{[I} X^J X^K X^{L]}\,.
\end{align}
\end{subequations}
\caption{``Standard'' charges densities of the M(atrix) model in a
  pp-wave, as derived in~\cite{Hyun:2002cm}. The charges are obtained
  by tracing over the~SU($N$) indices, i.e.~$Q_\alpha = \Tr
  q_\alpha$. For brevity we have suppressed an exponential involving
  the time coordinate; see (7)--(9) of~\cite{Hyun:2002cm}. This
  exponential is related to the time-dependence of the Killing
  spinors, which enters crucially in the construction of the
  supersymmetry transformation rules~\cite{Bain:2002tq}. Note that the
  anti-commutator in the first line is an anti-commutator of SU($N$)
  matrices, not a Dirac bracket.}
\label{t:Hcharges}
\end{table}
The algebra can be determined by systematically applying the Dirac
brackets
\begin{equation}
\big\{ (X^I)_a{}^b, (P^J)_c{}^d \big\}_{\text{DB}} = \delta^{IJ} \delta_a{}^d
\delta_c{}^b\, ,\quad
\big\{ (\theta^\alpha)_a{}^b, (\theta^\beta)_c{}^d \big\}_{\text{DB}} =
-i \delta^{\alpha\beta} \delta_a{}^d \delta_c{}^b\, .
\end{equation}
Here the indices $a,b,\ldots$ are in the fundamental of~SU($N$).  We will
  keep the subscript ``DB'' on the Dirac brackets to distinguish them
  from SU($N$) anti-commutators. The bosonic brane charges appear in
  the various brackets of the two supercharges~$Q$ and~$\tilde Q$, as
  computed by Hyun and Shin~\cite{Hyun:2002cm} (again, see
  footnote~\ref{f:conventions} for our conventions and
  table~\ref{t:Hcharges} for the explicit form of the generators),
\begin{equation}
\begin{aligned}
\big\{ q_{(\alpha\,a}{}^b, Q_{\beta)} \big\}_{\text{DB}} 
  = & -4i R\, {\cal H}_a{}^b \delta_{\alpha\beta}
   +i \frac{2\mu}{3} (\Pi \gamma^{ij} )_{\alpha\beta} (j^{ij})_a{}^b
   -i \frac{\mu}{3} (\Pi \gamma^{i'j'})_{\alpha\beta} (j^{i'j'})_a{}^b\\[1ex]
  & -2i\, \gamma^I_{\alpha\beta} (z^I)_a{}^b
   -2i\, \gamma^{IJKL}_{\alpha\beta} (z^{IJKL})_a{}^b\\[1ex]
 &  - \frac{\mu}{3} (\Pi \gamma^{j'})_{\alpha\beta}
     \big[ 2(X^i)^2 - (X^{i'})^2, X^{j'} \big]_a{}^b\\[1ex]
 & - \frac{\mu}{6} \epsilon_{ijk} \gamma^{iji'j'}_{\alpha\beta}
     \big[ X^{i'}, \{ X^k, X^{j'} \} \big]_a{}^b\,,\\[1ex]
\big\{ q_{\alpha\,a}{}^b, \tilde Q_{\beta}\big\}_{\text{DB}} 
  = & -2i\,\Big( P^I\gamma^I_{\alpha\beta} 
      - \gamma^{IJ}_{\alpha\beta} z^{IJ}
      -\frac{\mu}{3R} X^i (\Pi\gamma^i)_{\alpha\beta} 
      +\frac{\mu}{6R} X^{i'} (\Pi\gamma^{i'})_{\alpha\beta}
  \Big)_a^{\,\,\,b}\, .
\end{aligned}
\end{equation}
Here ${\cal H}_a{}^b$ denotes the Hamiltonian; we will not need its
explicit form but it can be found in~\cite{Hyun:2002cm} along with the
$\{ \tilde q, \tilde Q\}$ bracket (which we will also not need).  One
can verify, using a similar calculation, that the brackets of the
brane charges with themselves and with each other vanish identically
(this is true despite the fact that there is a momentum factor
appearing in the $z^I$~charge).
\medskip

Our main aim of this letter is to show that new fermionic brane
charges appear when one acts with a supersymmetry charge on the
bosonic brane charges. We will only show this for the two-form
charge~$Z^{IJ}$ as the story is very similar for the other brane
charges. By straightforward application of the basic Dirac brackets,
we find the key result
\begin{equation}
\label{e:QZ2}
\big\{ Q_{\alpha}, (z^{KL})_c{}^d \big\}_{\text{DB}} =
   (-i) \sqrt{\frac{R}{2}} \big[ X^K, (\gamma^L\theta)_\alpha
	  \big]_c{}^d 
   - (K\leftrightarrow L)\,,
\end{equation}
where $z^{KL}$ is the two-form brane charge density.  This calculation
shows that, indeed, the matrix model presents us with a new
\emph{fermionic brane charge}:
\begin{equation}
\label{e:newQ}
Q^{I}_\alpha := i\sqrt{\frac{R}{2}} \Tr\Big( 
  \big[X^I, \theta_\alpha \big] \Big)\, .
\end{equation}
We should emphasise that this calculation is completely identical to
the one in flat space. In a flat background, the new fermionic
operator~$Q^I_\alpha$ also follows from the algebra (as can be seen from
the fact that~\eqn{e:QZ2} does not depend on $\mu$). However, in flat
space one can (and typically does) consider representations for
which~\mbox{$Q^I_\alpha |\psi\rangle \equiv 0$}. The remaining
issue is therefore to show that such representations are
\emph{impossible} in the pp-wave background, because they would
violate the Jacobi identity $(Q,Q,Z)|\psi\rangle = 0$.
\medskip

This crucial Jacobi identity takes, in matrix model variables, the
more explicit form
\begin{equation}
\label{e:QQZmatrix}
0 =   \Big\{ \big\{ Q_{\alpha}, Q_{\beta} \big\}_{\text{DB}}, (z^{KL})_c{}^d \Big\}_{\text{DB}}
  -2\,\Big\{ \big\{ Q_{(\alpha}, (z^{KL})_c{}^d \big\}_{\text{DB}}, Q_{\beta)} \Big\}_{\text{DB}} \, .
\end{equation}
We first compute the intermediate result
\begin{subequations}
\begin{align}
\big\{ J^{ij}, (z^{KL})_{c}{}^d \big\}_{\text{DB}} &=
  2\, (z^{iK})_c{}^d \delta^{Lj} - 2\, (z^{iL})_c{}^d \delta^{Kj}\,,\\[1ex]
\big\{ J^{i'j'}, (z^{KL})_{c}{}^d \big\}_{\text{DB}} &=
  2\, (z^{i'K})_c{}^d \delta^{Lj'} - 2\, (z^{i'L})_c{}^d \delta^{Kj'}\,,
\end{align}
\end{subequations}
where anti-symmetry with unit weight in $i,j$ and $i',j'$ is implicitly
assumed on the right-hand side. Using this result we can compute the
first term in~\eqn{e:QQZmatrix}. One obtains a ``rotated'' bosonic
brane charge, simply because this charge carries space-time vector
indices:
\begin{multline}
\label{e:QQZ1}
\Big\{ \big\{ Q_{(\alpha}, Q_{\beta)} \big\}_{\text{DB}},
(z^{KL})_c{}^d\Big\}_{\text{DB}} \\[1ex]
=
  i\frac{2\mu}{3}\Big( 2\, (\Pi\gamma^{iL})_{\alpha\beta}
       (z^{iK})_c{}^d 
  - (\Pi\gamma^{i'L})_{\alpha\beta}
       (z^{i'K})_c{}^d\Big) - (K\leftrightarrow L)\, .
\end{multline}
Both sides are non-trivial when acting on a state $|\psi\rangle$ which
carries the bosonic brane charge.  Using~\eqn{e:QZ2} as well as the
symmetry properties of the gamma matrices as listed in
footnote~\ref{f:conventions}, the second term in~\eqn{e:QQZmatrix}
(including the ``$-2$'') is found to be
\begin{multline}
\label{e:QoldQnew}
-\big\{ Q_{\alpha}, (\gamma^K q^{L})_{\beta c}{}^d \big\}_{\text{DB}} - (\alpha\leftrightarrow\beta)\\[1ex]
  = (-i)\frac{2\mu}{3} \Big( 2\,(z^{iK})_c{}^d (\Pi\gamma_{i}{}^{L})_{\alpha\beta}
    -  (z^{i'K})_c{}^d (\Pi\gamma_{i'}{}^{L})_{\alpha\beta} \Big) -
  (K\leftrightarrow L)\, .
\end{multline}
Here $(q_{\alpha}^I)_a{}^b$ denotes the charge density of the new
fermionic brane charge~\eqn{e:newQ}. The \mbox{$\mu$-independent}
terms in this bracket are double commutators or commutators involving
the momentum variable, which should be set to zero.

The crucial point is now that in a pp-wave both sides of
equation~\eqn{e:QQZ1} act non-trivially on any state~$|\psi\rangle$
which carries the bosonic brane charge. Hence, in order to satisfy the
Jacobi identity, one has to make both sides of~\eqn{e:QoldQnew}
non-vanishing as well. That is, the new fermionic charge has to act
non-trivially on the state $|\psi\rangle$
(i.e.~$q^I_\alpha|\psi\rangle\not=0$). In that case we find that the
sum of~\eqn{e:QQZ1} and~\eqn{e:QoldQnew}, when acting on
$|\psi\rangle$, indeed vanishes. Contrast this with the situation in
flat space, where the right-hand side of~\eqn{e:QQZ1}
and~\eqn{e:QoldQnew} are zero because~$\mu=0$. In this case it is
consistent with the Jacobi identities to have a state with
non-vanishing $z^{KL}$ charge but vanishing $q^{I}_{\alpha}$
charge. These are indeed the representations which one usually
considers in flat space.

Summarising, we have shown that the operator algebra of the matrix
model in the pp-wave contains the new fermionic brane
charge~\eqn{e:newQ}, and that no representations exist for which
$Z^{IJ} |\psi\rangle\not=0$ but $Q^I_\alpha|\psi\rangle=0$.  Along
similar lines one can construct fermionic partners of the other
bosonic brane charges in table~\ref{t:Hcharges}. Details will appear
elsewhere.\footnote{One might expect, by very similar logic, that a
new supercharge is also required in order to satisfy the $(Q,Q,P)$
Jacobi identity. This situation is, however, slightly different. The
bracket of the supercharge with the momentum generator produces
\begin{equation}
\label{e:QPisnormalQ}
\big\{ Q_\alpha, P^I\big\}_{\text{DB}}
  = \sqrt{\frac{R}{2}} \frac{\mu}{6R} 
      \Big( 2\,(\Pi\gamma^i\theta)_\alpha 
              -(\Pi\gamma^{i'}\theta)_\alpha \Big)
  = \frac{\mu}{12} 
      \Big( 2\,(\Pi\gamma^i \tilde Q)_\alpha 
              -(\Pi\gamma^{i'} \tilde Q)_\alpha \Big)\, .
\end{equation}
The right-hand side is thus proportional to the ``old'' kinematic
supercharge~$\tilde Q$, which is the trace of the expression given
in~\eqn{e:Qkin}, and we do not obtain a new fermionic charge.}

\subsection{The supermembrane analogy}
\label{s:membrane}

Everything computed in the previous section has a direct analogue in
the supermembrane model. To give just one example, consider for
instance the supersymmetry transformation of the two-brane charge, in
matrix form expressed in~\eqn{e:newQ}. Using the elementary Dirac
bracket (see also footnote~\ref{f:DB} below)
\begin{equation}
\big\{ X^I(\sigma), P^J(\sigma') \big\}_{\text{DB}} =
\delta^{(2)}(\sigma-\sigma')\delta^{IJ}\, ,
\end{equation}
this transformation is now given by
\begin{multline}
\big\{Q_\alpha, Z^{KL}\big\}_{\text{DB}} 
= \bigg\{ \int_{\Sigma}\!{\rm d}^2\sigma\, P^I(\sigma) (\gamma_I\theta)_\alpha(\sigma)\,,\,\,
   \int_{\Sigma}\!{\rm d}^2\sigma'\, \epsilon^{rs}\, \partial_r X^K(\sigma')\, \partial_s X^L(\sigma')
\bigg\}_{\text{DB}} \\[1ex]
\begin{aligned}
&= -2\int_{\Sigma}\!{\rm d}^2\sigma\!\int_{\Sigma}\!{\rm d}^2\sigma'\,
  \big(\gamma^K\theta(\sigma) \big)_\alpha\,
    \epsilon^{rs} \left(\frac{\partial}{\partial \sigma'_r}
  \delta^{(2)}(\sigma-\sigma')\right)
   \partial_s X^L(\sigma') 
\\[1ex]
&= - 2\int_{\Sigma}\!{\rm d}^2\sigma\!\int_{\partial\Sigma}\!{\rm d}\sigma'\,
  n_r\epsilon^{rs}\,\big(\gamma^K\theta(\sigma) \big)_\alpha\,
      \delta^{(2)}(\sigma-\sigma')\, \partial_s X^L(\sigma') \\[1ex]
&= - 2\int_{\partial\Sigma}\! {\rm d}\sigma\,
     n_r\epsilon^{rs} \,
     \big(\gamma^K\theta(\sigma) \big)_\alpha\,
     \partial_s X^L(\sigma) \\[1ex]
&= -2\int\!{\rm d}^2\sigma\, \epsilon^{rs}
  \big(\gamma^K\partial_r\theta(\sigma) \big)_\alpha\,
    \partial_s X^L(\sigma) = 2 \gamma^K Q^L_\alpha \, .
\end{aligned}
\end{multline}
Here $n_r$ denotes the vector normal to the integration boundary, and
anti-symmetry in the $K,L$ indices is again implicitly understood
everywhere.  This calculation\footnote{\label{f:DB} Strictly speaking,
we have here used a Dirac bracket which does \emph{not} preserve the
boundary conditions. The true Dirac brackets, which incorporate the
boundary conditions by treating them as constraints, lead to a
dynamical evolution in which the winding charges are not dynamical
variables~\cite{Sheikh-Jabbari:1999xd}. What we have computed here is
the world-volume version of the ``spectrum generating'' algebra, which
relates physical states with different boundary conditions. Compare
this with e.g.~the action of rotation generators on bosonic winding
charges,
\begin{equation}
\big\{ M_{IJ}, Z^{KL} \big\} = 4\,\delta_{[I}{}^{[K}  Z_{J]}{}^{L]}\, .
\end{equation}
This action changes the boundary conditions and produces new
configurations which are not related to the old ones by
dynamical evolution.}  shows that the new fermionic
charge~$Q^I_\alpha$ is non-zero whenever~$\theta(\sigma)$ is not
single-valued on the membrane world-surface, or for open membranes,
whenever $\theta(\sigma)$ takes different values at the two boundaries
of the membrane.\footnote{A similar calculation for the string in flat
space-time was done by Hatsuda and
Sakaguchi~\cite{Hatsuda:2000hp}. For historical completeness, we
should also mention that a fermionic extension of the superalgebra of
the string in an anti-de-Sitter background was considered by
Hatsuda~\cite{Hatsuda:2000mn}. However, this paper constructs the new
fermionic charge as the superpartner of the momentum generator (just
like in Green's original construction~\cite{Green:1989nn}) and does
not take into account winding charges. Another related paper
is~\cite{Hatsuda:2003er}, which only considers particle world-line
superalgebras and therefore misses the central charges as well.}

In the pp-wave background, such configurations with
non-trivial~$\theta(\sigma)$ behaviour indeed do exist! They are
simplest to analyse for open strings, whose algebra can be shown to
contain a topological fermionic charge similar to~\eqn{e:newQ}:
\begin{equation}
\label{e:Qnewstring}
Q_w = \int_0^\pi\!{\rm d}\sigma\, (1-P) \partial_\sigma \theta_1\, .
\end{equation}
Here $P$ is the matrix which relates the two fermions $\theta_1$ and
$\theta_2$ of the open string, implementing the boundary
conditions. In a pp-wave background, the mode expansions of the
fermions typically contain zero-modes which are independent of the
world-sheet time~$\tau$ but do depend on~$\sigma$. The zero modes for
a string with D1-brane boundary conditions are, for instance, given
by~\cite{Bain:2002tq}
\begin{equation}
\begin{aligned}
\theta^1 &= (1+\Gamma^{+-} \Pi)\, \theta^+ e^{\mu \sigma} + 
           (1- \Gamma^{+-} \Pi)\, \theta^- e^{-\mu \sigma} \,,\\[1ex]
\theta^2 &= (\Gamma^{+-} + \Pi)\, \theta^+ e^{\mu \sigma}\;\; + 
           (\Gamma^{+-} - \Pi)\, \theta^- e^{-\mu \sigma}\,,
\end{aligned}
\end{equation}
for arbitrary constant spinors $\theta^+$, $\theta^-$. When inserted
in~\eqn{e:Qnewstring}, these zero modes are responsible for a
non-vanishing fermionic topological charge of the open
string. Moreover, the fermionic zero modes are related, by a simple
supersymmetry transformation, to bosonic zero-modes, which also depend
non-trivially on~$\sigma$ (again, as explained in footnote~\ref{f:DB},
one acts with a broken supersymmetry transformation and thereby
changes the boundary conditions). This analysis, which crucially
differs from flat space because no zero-modes for the fermions exist
in that case, can be extended to the supermembrane in a
straightforward way.

\section{Discussion and conclusions}
\label{s:covariant}

We have shown that new fermionic brane or winding charges appear in
the supermatrix model as well as the supermembrane. The presence of
these charges was expected from a Jacobi-identity argument, but we
have shown here that such charges are indeed present in explicit
models. We have also shown that physical multiplets involving bosonic
branes will always also contain states which carry these new fermionic
brane charges.

This observation has important consequences for the construction of
extensions of superisometry algebras of maximally supersymmetric
backgrounds.  In~\cite{Meessen:2003yi} we have shown that the
$\text{AdS}_7\times S^4$ superisometry algebra $\ospsixtwo$ cannot be
extended (in a physically acceptable way) with only bosonic brane
charges. Taking into account additional fermionic brane charges leads
to an algebra which resembles an ``anti-de-Sitter version'' of
Sezgin's \mbox{M-algebra}~\cite{Sezgin:1997cj}. One should note,
however, that the algebras presented in the present letter are based
solely on the non-trivial bracket $[Q,Z]=Q'$, while Sezgin's proposal,
following Green~\cite{Green:1989nn}, has in addition $[Q,P]=Q''$. In
the supermatrix model or the supermembrane, the bracket $[Q,P]$ does
not lead to a new supercharge, so it is unclear whether the M-algebra
is indeed only consistent upon introduction of a new superpartner for
the momentum generator. A more elaborate analysis of this problem will
appear elsewhere.

It is clearly necessary to develop a better understanding of branes
which carry the new fermionic charges and the supermultiplets in which
they fit. One would perhaps also like to understand them in terms of
supergravity solutions. Finally, we should mention that it would be
very interesting to understand the implications of these new fermionic
brane charges in the context of the AdS/CFT correspondence.

\section*{Acknowledgements}

We thank Eric Bergshoeff, Ergin Sezgin, Paul Townsend and especially
Bernard de Wit for comments on a preliminary version of this letter.

\bibliographystyle{JHEP}
%\bibliography{kasbib}
\bibliography{matrix_charges}
\end{document}